\documentclass[aps,preprint,groupedaddress]{revtex4}

\usepackage{graphicx}

\newcommand{\be}{\begin{equation}}
\newcommand{\ee}{\end{equation}}
\newcommand{\bea}{\begin{eqnarray}}
\newcommand{\eea}{\end{eqnarray}}

\begin{document}
\bibliographystyle{apsrev}

 \preprint{UAB-FT-489}

\title{Neutrino-photon reactions for energies above $m_e$}



\author{Eduard Mass{\'o}} 
\email[]{masso@ifae.es}
\author{Francesc Rota}
\email[]{rota@ifae.es}

\affiliation{Grup de F{\'\i}sica Te{\`o}rica and Institut 
de F{\'\i}sica d'Altes
Energies\\Universitat Aut{\`o}noma de Barcelona\\ 
08193 Bellaterra, Barcelona, Spain}


\date{\today}

\begin{abstract}
We show that neutrino-photon reactions above $m_e$ are dominated by
the reaction $\nu \gamma \rightarrow \nu  e^+ e^-$. We calculate
its cross-section 
and see that it is larger by several orders of magnitude than
the cross-sections of other neutrino-photon processes,
for energies above $m_e$. 
We also discuss potential astrophysical and cosmological consequences.
\end{abstract}

\maketitle


Neutrino-photon reactions could play a role in some astrophysical
or cosmological environments. Recently, there has been some interest in 
these type of interactions. (See Refs.  
\cite{earlynp}-\cite{aghababaie}.) Let us consider
\be\label{X}
\nu\, \gamma\ \rightarrow\ \nu\, X
\ee 
This reaction has to involve weak interactions since
the neutrino has no electric charge and the expected magnetic
moment is small. The elastic channel
\be\label{elastic}
\nu\, \gamma\ \rightarrow\ \nu\, \gamma
\ee is further suppressed due to the prohibition of a  two-photon
coupling to a $J=1$ state (Yang theorem). The
cross-section is on the order of 
$\sigma \sim G_F^2 \alpha^2 \omega^6/M_W^4$
\cite{earlynp,dicus1}
where $\omega$ is the energy of the photon in the center-of-mass system 
($\sqrt s = 2 \omega$) and where we assume we are in the limit 
$\omega<<M_W$ and also $m_\nu<<\omega$.  The Yang suppression amounts to 
the small factor $(\omega/M_W)^4$ in $\sigma$. 
We refer the reader to \cite{dicus1}
where the cross-section of the process (\ref{elastic}) in the standard model 
has been calculated numerically.

Recently, in a series of papers
\cite{dicus2}-\cite{abada2}
it has been realized that the inelastic process
\be\label{inelastic}
\nu\, \gamma\ \rightarrow\ \nu\, \gamma\, \gamma
\ee 
largely dominates over (\ref{elastic}). 
The technicalities for the calculation of the cross-section
are different depending on whether we have $\omega<<m_e$ or
not.

For $\omega<<m_e$, one has a cross-section of magnitude  
$\sigma \sim G_F^2 \alpha^3 \omega^{10}/m_e^8$
\cite{dicus2,abada1}. 
Now there is no
Yang suppression, as can be readily seen in the expression for the
cross-section, and instead we have an inverse power of the electron
mass. Indeed, (\ref{inelastic}) is related to  
$\gamma \gamma \rightarrow \gamma \gamma$ scattering when
substituting a photon by the neutrino current, and for
$\omega<<m_e$ one can make use of the effective Euler-Heisenberg 
lagrangian that has the inverse power $m_e^{-4}$ of the scale $m_e$.
This simplifies the calculation and explains the appearance of the
electron mass in the cross-section.
The change of the $M_W$ scale by the $m_e$ scale explains the enhancement
of the inelastic channel (\ref{inelastic}).  

When $\omega>m_e$, the cross-section has to be calculated
directly and there is no simplification of the type we have
mentioned. The calculation (for $\omega<<M_W$) has been carried in 
\cite{dicus3,abada2}.
From a theoretical point of view, this calculation is
interesting since it is a nice example where one can compare the
results using an effective lagrangian and the results of a full
calculation that can be used at high energies. The authors argue 
that the region $\omega>m_e$ may have some phenomenological interest. 
For example, if one wants to check whether the reaction
(\ref{inelastic}) contributes to the neutrino opacities in the
first stages of a supernova  one needs clearly the cross-section
when  $\omega>m_e$. Also, the authors of \cite{dicus3}
as well as the authors of \cite{abada2}
consider some possible cosmological applications. They notice
that, if there is any of such applications, it would occur in
the regime $\omega>m_e$.

We would like to point out that when one crosses the $m_e$
threshold the reaction
\be\label{ours}
\nu\, \gamma\ \rightarrow\ \nu\, e^+\, e^-
\ee
dominates over the reaction (\ref{inelastic}). In fact, the cross-sections
are larger by several orders of magnitude than the cross-sections of
(\ref{inelastic}).
The reason is simple to understand if one considers the limit 
$M_W>>\omega>>m_e$. In this limit,
while (\ref{inelastic}) is on the order of $G_F^2 \alpha^3 \omega^2$,
(\ref{ours}) is of order $G_F^2 \alpha \omega^2$.  Due to
the potential interest of the process, we would like to 
present its calculation for all energies $\omega$ (below $M_W$).

First of all, we show the associated Feynman diagrams for the process 
in figure 1. We assume the limit $\omega<<M_W$, so that we can use
the Fermi effective coupling,
\be\label{fermi}
{\cal L}_F = - {4 G_F\over\sqrt 2}\ \bar \Psi_\nu \gamma_\mu P_L \Psi_\nu\
\bar \Psi_e \gamma^\mu (c_LP_L + c_R P_R) \Psi_e
\ee
with $P_L=(1-\gamma_5)/2$ and $P_R=(1+\gamma_5)/2$.
When $\nu=\nu_\mu$ or $\nu=\nu_\tau$ there is only the neutral
weak current contributing to (\ref{fermi}), and we have
\bea 
c_L &=& - 1 / 2 + \sin^2\theta_W   \cr
c_R &=& \sin^2\theta_W  \label{nu_mu}
\eea
($\theta_W$ is the weak mixing angle). When $\nu=\nu_e$, apart from
the neutral current there is also the charged current contribution. It
can be Fierz rearranged to have the form of (\ref{fermi}), and then
\bea \label{nu_e}
c_L &=&  1 / 2 + \sin^2\theta_W  \cr
c_R &=& \sin^2\theta_W 
\eea

Next, we write the amplitude for 
$\nu(p)+ \gamma(k) \rightarrow \nu(q)+ e^+(q_1) + e^-(q_2)$
\bea\label{amplitude}
& M &=  - {4 i e G_F \over \sqrt 2}\  \epsilon_\mu(k)\
 \bar u(q) \gamma_\nu P_L u(p)\ \times  \cr  
&& \bar u(q_1) \left[\gamma^\mu 
{1\over {/\kern-0.5em q_1} - {/\kern-0.5em k} - m} 
\gamma^\nu (c_LP_L + c_RP_R)  
 +  \gamma^\nu (c_LP_L + c_RP_R)
{1 \over - {/\kern-0.5em q_2} + {/\kern-0.5em k} -m}\gamma^\mu \right] v(q_2)
\eea

Finally, we evaluate numerically the cross-section. 
We work, as we said, in the limit $\omega << M_W$, as well as 
$m_\nu<<\omega$.
The results, for the case $\nu=\nu_e$ are presented in figure 2, in
the range $m_e < \omega < 10^2 m_e$. For higher energies, the 
cross-section scales approximately as $\omega^2$. For instance, 
for $\omega=100$ MeV,
we have $\sigma \simeq 2\times 10^{-3}$ fb, and for $\omega=1$ GeV,
it is already $\sigma \simeq 0.4$ fb.
The cross-section for our process (\ref{ours}) when $\nu$ is
either $\nu_\mu$ or $\nu_\tau$ is obtained when using (\ref{nu_mu})
instead of (\ref{nu_e}). As expected, one gets numerically a similar 
result.

In figure 2 we also show the cross-section
for the reaction (\ref{inelastic}), and as expected we see that it
is smaller by several orders of magnitude. Of course, our reaction
has a threshold at $\omega=m_e$. For energies below $m_e$, the dominant
process is (\ref{inelastic}).

Let us now examine potential consequences in the supernova
dynamics.
Any reaction of the type (\ref{X}) contributes to the neutrino opacity
in a supernova collapse. In the conservative range of temperatures 
$T=10-100$ MeV,
both photons and neutrinos have energies exceeding $m_e$ and thus
it is our reaction (\ref{ours}) that will be most important. 
To evaluate its possible role, we need the neutrino mean free
path,
\be
\lambda = {1 \over \sigma  n_\gamma}
\ee
due to (\ref{ours}).
We estimate $\sigma$ by evaluating it at the average energy
$\omega \sim 3T$. In the temperature range we have indicated,
we get
\be\label{figures}
\lambda \sim 3 \times 10^8\ -\  2 \times 10^3\ \ {\rm cm}
\ee
(the shorter $\lambda$ corresponding to the higher temperature, of
course).

Neutrino scattering with non-relativistic nucleons
  in a supernova has a cross-section
$\sigma \sim 2 \times 10^{-5} (E/{\rm MeV})^2$ fb, where $E$ is the neutrino 
energy. The large nucleon density, $n_N \sim 2 \times 10^{38}$
cm$^{-3}$, leads to $\lambda =1/\sigma n_N \sim  3-300$ cm. 
At the view of the figures in (\ref{figures}) 
we tentatively conclude that the role of neutrino-photon reactions in the 
neutrino opacity in the supernova is small.

Regarding cosmological applications, the interest of reactions (\ref{X}) 
in the early universe has been discussed in Refs. \cite{dicus4} and  
\cite{abada2} (See also Ref.\cite{harris}.) 
In the standard scenario, we have that for temperatures 
$T> 1$ MeV, weak interactions have not yet decoupled and neutrinos interact 
with matter, in particular with electrons. Electrons in turn interact
with photons. Thus, neutrinos, photons and electrons are in equilibrium.
At $T\sim 1$ MeV, neutrinos decouple.
  
Once we have direct neutrino-photon interactions, it may be
interesting to investigate at which temperature $T_D$ these 
direct interactions decouple. 
In \cite{dicus4}
it is shown that, due to the reaction (\ref{elastic}), neutrinos decouple
from photons at $T_D \sim 1$ GeV. In \cite{abada2}, the
authors consider the inelastic process  (\ref{inelastic})
and demonstrate that $T_D \sim 1$ GeV. We expect a lower 
decoupling temperature due to our reaction (\ref{ours}). Let us show
that this is the case.

We have, on the one hand that the expansion rate of the Universe is given by
\be
H = 1.66\ g^{1/2}\ {T^2 \over M_{Pl}}
\ee
where $M_{Pl}$ is the Plank energy. Anticipating 
that   1 MeV $ <T_D< $ 100 MeV, we set the total degrees of freedom
$g=10.75$. 
On the other hand, the interaction rate is given by
\be
\Gamma  = \sigma  n_\gamma
\ee
Again, to estimate $\sigma$, we use the average energy $\omega \sim
3 T$. At high energies, one has that the ratio $ \Gamma/H >> 1$,
 as expected. 
This ratio decreases with decreasing temperature and becomes of order
unity for $T=T_D \sim 10$ MeV. This is the decoupling temperature
of neutrinos and photons due to the process (\ref{ours}), and we see 
that is below the one found using other neutrino-photon reactions. 
In any case, neutrinos are not decoupled due to weak interactions
until $T_D \sim 1$ MeV, and we are not able to see any interesting
application of our result.


\begin{acknowledgments}
Work partially supported by the CICYT Research Project
AEN99-0766. We would like to thank Llu{\'\i}s Ametller, Fernando Cornet
and Jaume Guasch for helpful discussions.

\end{acknowledgments}


\begin{thebibliography}{99}

\bibitem{earlynp}
H.-Y Chiu and P. Morrison,
Phys.\ Rev.\  Lett.\ {\bf 5}, 573 (1960) \\
V.~K.~Cung and M.~Yoshimura,
Nuovo Cim.\  {\bf 29A} 557  (1975)

\bibitem{dicus1}
D.~A.~Dicus and W.~W.~Repko,
Phys.\ Rev.\  {\bf D48}, 5106 (1993)

\bibitem{dicus2}
D.~A.~Dicus and W.~W.~Repko,
Phys.\ Rev.\ Lett.\  {\bf 79} 569 (1997) 

\bibitem{dicus3}
D.~A.~Dicus, C.~Kao and W.~W.~Repko,
Phys.\ Rev.\  {\bf D59} 013005 (1999) 

\bibitem{abada1}
A.~Abada, J.~Matias and R.~Pittau,
Phys.\ Rev.\  {\bf D59} 013008 (1999)

\bibitem{abada2}
A.~Abada, J.~Matias and R.~Pittau,
Nucl.\ Phys.\  {\bf B543} 255 (1999)

\bibitem{dicus4}
A.~Abbasabadi, A.~Devoto, D.~A.~Dicus and 
W.~W.~Repko,
Phys.\ Rev.\  {\bf D59}  013012 (1999) 

\bibitem{harris}
M.~Harris, J.~Wang and V.~L.~Teplitz,
astro-ph/9707113.

\bibitem{aghababaie}
Y.~Aghababaie and C.~P.~Burgess,
hep-ph/0006165.

\end{thebibliography}

\newpage

%
%
 \begin{figure}
 \label{figure1}

\begin{center} 

 \includegraphics[width=13cm]{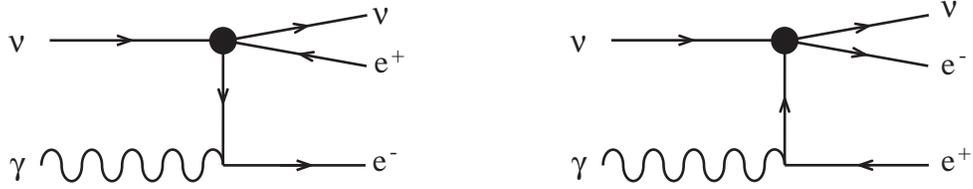}

\end{center}

 \caption{Feynman diagrams for the process 
$\nu \gamma \rightarrow \nu e^+ e^-$.}

 \end{figure}

 \begin{figure}
 \label{figure2}

\begin{center} 

 \includegraphics[width=13cm]{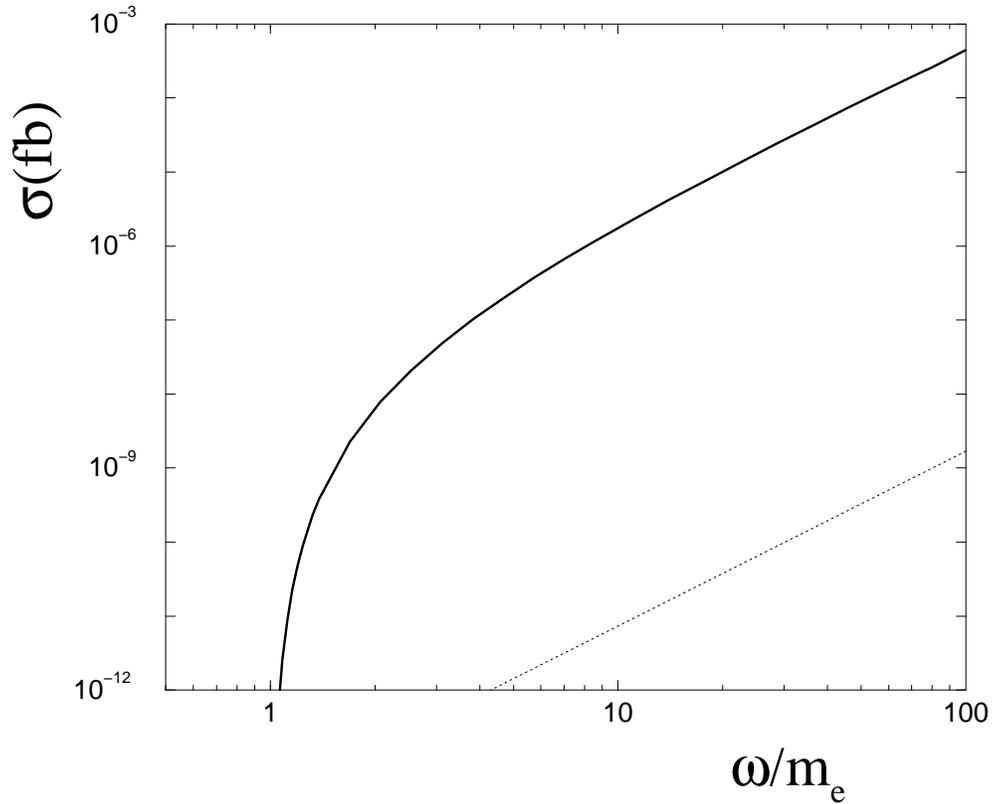}

\end{center}

 \caption{ The cross-section for the process 
$\nu \gamma \rightarrow \nu e^+ e^-$ in fb as a function of $\omega/m_e$ with
  $\omega$ the center-of-mass energy of the photon. We display
the cross-section for the process  
$\nu \gamma \rightarrow \nu \gamma \gamma$ (dotted line).}
 \end{figure}

%
%

\end{document}